\documentclass[11pt]{article}
\usepackage{geometry}  
\usepackage{graphicx}
\usepackage{amssymb}
\usepackage{amsmath}
\usepackage{epstopdf}
\usepackage{hyperref}
\usepackage{url}

\def\R{{\rm I\!R}}
\def\alphabar{\bar{\alpha}}
\def\betabar{\bar{\beta}}

\def\ie {{i.e.}}

\def\EQUIV {\Leftrightarrow}

\def\calY {{\cal Y}}
\def\calV {{\cal V}}
\def\Gs {\Gamma ^*}

\title{Orbifold construction of the modes\\
  of the Poincar\'e dodecahedral space}
\author{\begin{tabular}{ccc}
Marc Lachi\`eze-Rey &and& Jeffrey Weeks\\
APC
&& 15 Farmer Street\\
(Astroparticule et Cosmologie)
&& Canton NY\\
CNRS-UMR 7164, France&& USA
\end{tabular}}

\begin{document}
\maketitle

\begin{abstract}
  \noindent We provide a new construction of the modes of the
  Poincar\'e dodecahedral space $S^3/I^*$.  The construction
  uses the Hopf map, Maxwell's multipole vectors and
  orbifolds.  In particular, the *235-orbifold serves
  as a parameter space for the modes of $S^3/I^*$,
  shedding new light on the geometrical significance
  of the dimension of each space of $k$-modes, as well
  as on the modes themselves.
  \\

\end{abstract}

\section{Introduction}
\label{SectionIntroduction}
%

Cosmological motivations \cite{LaLu} have inspired recent progress
in understanding the eigenmodes of the spherical spaces
$S^3/\Gamma^*$, \ie, the quotients of the three-sphere $S^3$ by a
binary polyhedral group $\Gamma^*$. Such modes may be seen as the
$\Gamma ^*$-invariant solutions of the Helmoltz equation in the
universal cover $S^3$. Their numeration and degeneracy were given
by Ikeda \cite{Ikeda}. Recent works
\cite{Caillerie,Gundermann,mlrEigenmodes} have provided  various
means to calculate them.

Here we give a new point of view, using the Hopf map, multipole
vectors and orbifolds to construct the modes of $S^3/\Gamma^*$ and
shed additional light on the geometrical significance of Ikeda's
formula. Section~\ref{SectionHopf} reviews the Hopf map and uses
it to lift eigenmodes from $S^2$ to $S^3$.
Section~\ref{SectionEigenmodes} uses twist operators to extend the
lifted modes to a full eigenbasis for $S^3$.
Section~\ref{SectionModesOfSphericalSpaces} generalizes the
preceding results from the modes of $S^3$ to the modes of a
spherical space $S^3/\Gs$, showing that the latter all come from
the lifts of the those eigenmodes of $S^2$ that are invariant
under the corresponding (non-binary) polyhedral group $\Gamma$. We
then turn to a detailed study of the $\Gamma$-invariant modes of
$S^2$. Section~\ref{SectionS2Modes} recalls Maxwell's multipole
vector approach and uses it to associate each mode of $S^2/\Gamma$
to a $\Gamma$-invariant set of multipole directions.  Restricting
attention to the case that $\Gamma$ is the icosahedral group,
Section~\ref{SectionPDSModes} introduces the concept of an
orbifold and re-interprets a $\Gamma$-invariant set of multipole
directions as a (much smaller) set of points in the *235-orbifold,
which serves as the parameter space.
Section~\ref{SectionConclusion} pulls together the results of the
preceding sections to summarize the construction of the modes of
the Poincar\'e dodecahedral space and state the dimension of the
mode space for each $k$.

%
\section{From $S^2$ to $S^3$: lifting with the Hopf map}
\label{SectionHopf}
%

\subsection{Spheres}
\label{SectionSpheres}

We parameterize the circle $S^1$ as the set of points $\alpha \in
\mathbb C$ of unit norm $\alpha\bar\alpha = 1$. The relationship
between the complex coordinate $\alpha$ and the usual Cartesian
coordinates $(x,y)$ is the natural one: $\alpha = x + i y$.

We parameterize the 2-sphere $S^2$ as the set of points $(x,y,z)
\in \mathbb R^3$ of unit norm $x^2 + y^2 + z^2 = 1$.

We parameterize the 3-sphere $S^3$ as the unit sphere in $\mathbb
C^2$:   the set of points $(\alpha, \beta)
\in \mathbb C^2$ of unit norm $\alpha\bar\alpha + \beta\bar\beta =
1$.  Hereafter, we will always assume that this normalization relation holds. The relationship between the complex coordinates $(\alpha,
\beta)$ and the usual Cartesian coordinates $(x,y,z,w)$ is the
natural one: $\alpha = x + i y$ and $\beta = z + i w$.
\\

\begin{figure}[!ht]
\begin{center}
\includegraphics[width=8cm]{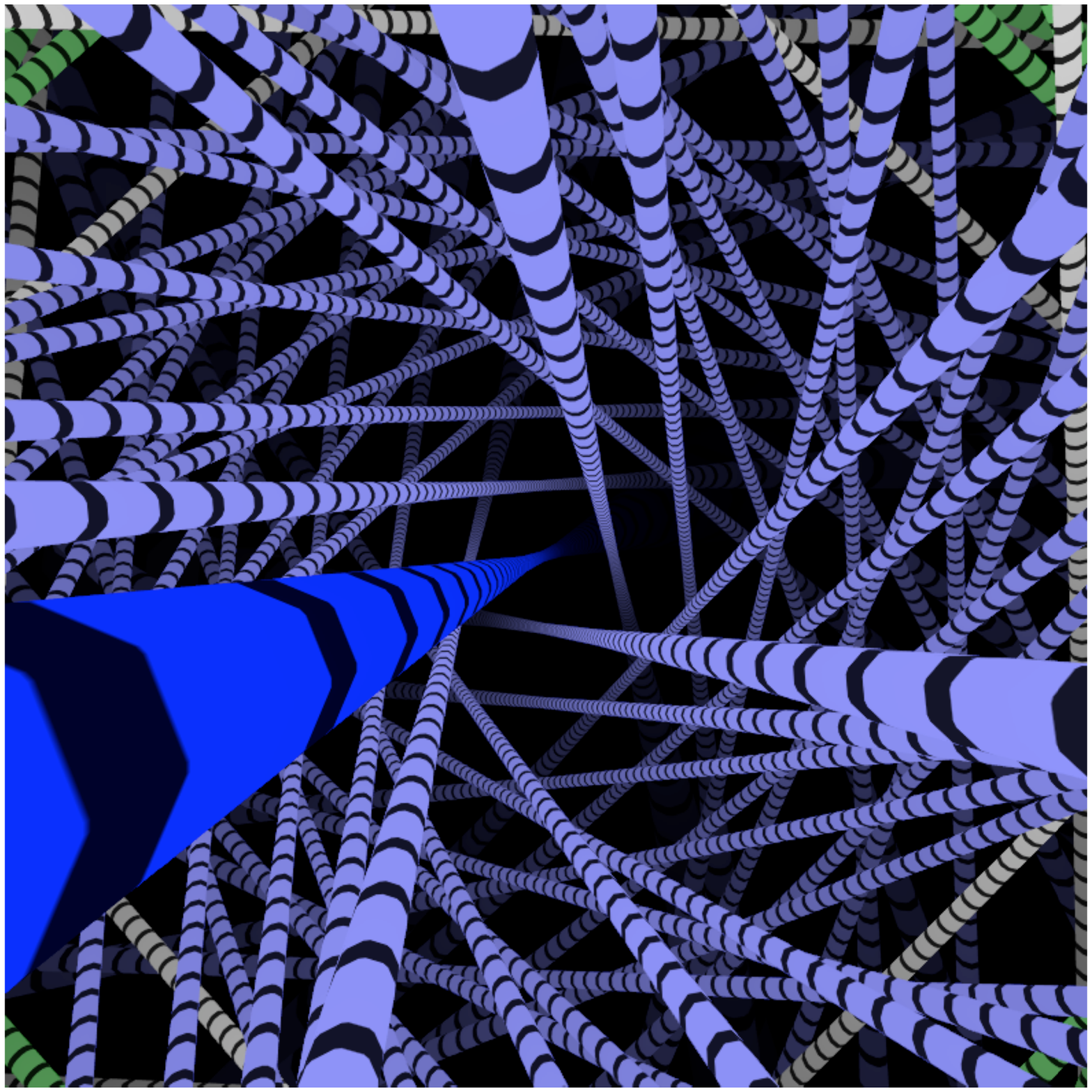}
\caption{A computer generated view of the Clifford parallels in
$S^3$} \label{FigureCliffordParallels}
\end{center}
\end{figure}

\subsection{The Hopf fibration}
\label{SectionHopfFibration}

In $S^3$, simultaneous rotation in the $\alpha$- and $\beta$-planes defines
the {\it Hopf flow} $H_t: S^3 \rightarrow S^3$,
\begin{equation}
\label{HopfFlow}
  H_t(\alpha, \beta) \equiv
   (\;e^{i t}\alpha,\;e^{i t}\beta\;).
\end{equation}
The Hopf flow is homogeneous in the sense that it looks the same
at all points.  An orbit
\begin{equation}
\label{CliffordParallel}
  \{\;(e^{i t} \alpha, e^{i t} \beta)\quad|\quad 0 \leq t < 2\pi\}.
\end{equation}
is a  great circle  on $S^3$ called a {\it Clifford parallel}
(Figure~\ref{FigureCliffordParallels}). Collectively the Clifford
parallels comprise the {\it Hopf fibration} of $S^3$. The fibers
carry Clifford's name because William Kingdon Clifford (1845 --
1879) discovered them before Heinz Hopf (1894 -- 1971) was born.
However, while Clifford understood the fibration quite well, he
did not, as far as we know, go on to consider the quotient map
(Eqn.~(\ref{HopfMap})).

As we walk along any given Clifford parallel $(e^{i t} \alpha,
e^{i t} \beta)$, the ratio of its coordinates $\frac{e^{i t}
\alpha}{e^{i t} \beta}$ remains a constant $\frac{\alpha}{\beta}$,
independent of $t$.  The ratio $\frac{\alpha}{\beta}$ labels uniquely each  Clifford parallel,
taking values in the extended complex numbers $\mathbb C \cup
\{\infty\}$, where $\infty$ represents the ratio
$\frac{\alpha}{\beta} = \frac{1}{0}$.  The extended complex
numbers may be visualized as a Riemann sphere, proving that the
Clifford parallels are in one-to-one correspondence with the
points of a topological 2-sphere $S$.

The {\it Hopf map} is defined as sending any point
$(\alpha,\beta)$ of $S^3$ to the fiber its belong to, \ie, the
point of $S$ labelled by $\frac{\alpha}{\beta}$.  Composing with a
natural map from $S$ to the unit 2-sphere $S^2$ gives an explicit
formula for the Hopf map:
\begin{eqnarray}
\label{HopfMap}
  &p: S^3 \rightarrow S^2&\nonumber\\
 &(\alpha,\beta) \rightarrow
  p(\alpha,\beta) =(x,y,z)=
  \left(\;\alpha\bar\beta + \bar\alpha\beta,\;
       -i\;(\alpha\bar\beta - \bar\alpha\beta),\;
          \beta\bar\beta - \alpha\bar\alpha\;
        \right)&.
\end{eqnarray}
It is easy to check that $x^2+y^2+z^2=1$, confirming that the Hopf
map $p$ sends $S^3$ to the the unit 2-sphere.

\subsection{Lifts of functions}
\label{SectionLifts}

Any given function $f$ on $S^2$ lifts to a function $F$ on $S^3$
by composition with the Hopf map~$p$ from
Equation~(\ref{HopfMap}),
\begin{equation}
  F: S^3 \xrightarrow{\;p\;} S^2 \xrightarrow{\;f\;}\mathbb R.
\end{equation}
In other words, $F = p^*f$ is the pull-back of $f$ by $p$:  explicitly,
\begin{equation}
\label{LiftingFormula}
  F(\alpha, \beta) \equiv f(p(\alpha, \beta)) =
    f\left(\alpha\bar\beta + \bar\alpha\beta,\;
       -i\;(\alpha\bar\beta - \bar\alpha\beta),\;
          \beta\bar\beta - \alpha\bar\alpha
        \right).
\end{equation}
For example, the quadratic polynomial
\begin{equation}
  f(x,y,z) = x^2 - y^2,
\end{equation}
lifts to the quartic polynomial
\begin{eqnarray}
  F(\alpha, \beta)
    &=& (\alpha\bar\beta + \bar\alpha\beta)^2
      - (-i\;(\alpha\bar\beta - \bar\alpha\beta))^2\nonumber\\
    &=& 2(\alpha^2\bar\beta^2 + \bar\alpha^2\beta^2).
\end{eqnarray}

\noindent{\bf Definition \ref{SectionLifts}.1.}  We call a
function $F: S^3 \rightarrow \mathbb R$ {\it vertical} if it is
constant along every Clifford parallel
(Formula~(\ref{CliffordParallel})).
\\

\noindent For every function $f: S^2 \rightarrow \mathbb R$, the
construction of the lift $F(\alpha, \beta) = f(p(\alpha, \beta))$
guarantees that $F $ is vertical.
\\

\noindent{\bf Proposition \ref{SectionLifts}.2.}  The Hopf map
lifts a polynomial $f: S^2 \rightarrow \mathbb R$ of degree $\ell$
to a polynomial $F: S^3 \rightarrow \mathbb R$ of degree $2\ell$.
\\

\noindent{\it Proof.}  The lifting formula~(\ref{LiftingFormula})
doubles the degree of any polynomial.
$\blacksquare$\\

%
\section{Eigenmodes}
\label{SectionEigenmodes}
%

\subsection{Basic definitions}
\label{SubsectionEigenmodesDefinitions}

\noindent{\bf Definition \ref{SubsectionEigenmodesDefinitions}.1.}
An {\it $\ell$-eigenmode} is an eigenmode $f: S^2 \rightarrow
\mathbb R$ of the Laplacian, with eigenvalue $\lambda_\ell =
\ell(\ell + 1)$.
\\

\noindent An $\ell$-eigenmode is a solution of the Helmholtz
equation
\begin{equation}
\label{HelmoltzS2}
  \Delta_{S^2} f = \ell(\ell + 1) f.
\end{equation}
The index $\ell$ takes values in the set $\{0, 1, 2,\ldots\}$. For
each $\ell$, the $\ell$-eigenmodes (which are the usual spherical
harmonics) form a vector space $V^\ell$ of dimension $2\ell + 1$.\\

\noindent{\bf Definition \ref{SubsectionEigenmodesDefinitions}.2.}
A {\it $k$-eigenmode} is an eigenmode $F: S^3 \rightarrow \mathbb
R$ of the Laplacian with eigenvalue $\lambda_k = k(k + 2)$.
\\

\noindent A $k$-eigenmode is a solution of the Helmholtz equation
\begin{equation}
\label{HelmoltzS3}
  \Delta_{S^3}F = k(k + 2)~F.
\end{equation}
The index $k$ takes values in the set $\{0, 1, 2,\ldots\}$. For
each $k$, the $k$-eigenmodes form a vector space $V^k$ of
dimension $(k + 1)^2$.
\\

\subsection{Eigenmodes of $S^2$ define eigenmodes of $S^3$}
\label{SectionLiftingModes}

\noindent{\bf Proposition \ref{SectionLiftingModes}.1.}  An
$\ell$-eigenmode $f$  on the unit 2-sphere lifts to a
$k$-eigenmode $F$ on the unit 3-sphere, with $k = 2\ell$.
\\

\noindent{\it Proof.} It is well-known that the $\ell$-eigenmodes
are precisely the homogenous harmonic polynomials of degree $\ell$
on $\mathbb R^3$, with domain restricted to the unit 2-sphere.
Similarly the $k$-eigenmodes are the homogeneous harmonic
polynomials of degree $k$ on $\mathbb R^4$, with domain restricted
to the unit 3-sphere.  A harmonic function on $\R^4$ satisfies
\begin{equation}
  \Delta _{R^4} F \equiv
   4~(\partial_\alpha~\partial_{\alphabar}
   + \partial_\beta~\partial_{\betabar}~) F = 0.
\end{equation}
When $F$ is the pull-back of $f$ given by (\ref{LiftingFormula}),
direct calculations give
\begin{equation}
  \Delta _{R^4} F(\alpha,\beta) =
    (\partial_x~\partial_x +
     \partial_y~\partial_y +
     \partial_z~\partial_z) f(x,y,z)
     = \Delta_{R^3} f(x,y,z).
\end{equation}
Thus, the pull-back of a harmonic function on $\R^3$ is a harmonic
function on $\R^4$, and therefore the pull-back of an eigenmode of
$\Delta_{S^2}$ is an eigenmode of $\Delta_{S^3}$.  Together with
Proposition~\ref{SectionLifts}.2, this completes the proof.
$\blacksquare$\\

\noindent{\bf Notation \ref{SectionLiftingModes}.2.}  Let $Y_{\ell
m}$ denote the usual spherical harmonics on $S^2$. For example,
the $Y_{2,m}$ may be expressed as harmonic polynomials as follows
\begin{center}
\begin{tabular}{|l|l|l|}
\hline
  & trigonometric & polynomial \\
\hline
& & \\ $Y_{2,+2}$ & $\sqrt\frac{15}{32\pi}\;\sin^2\theta\;e^{2i\varphi}$ & $\sqrt\frac{15}{32\pi}\;(x + i y)^2$ \\
 $Y_{2,+1}$ & $\sqrt\frac{15}{8\pi}\;\cos\theta\;\sin\theta\;e^{i\varphi}$ & $\sqrt\frac{15}{8\pi}\;z(x + i y)$ \\
 $Y_{2,0} $ & $\sqrt\frac{5}{16\pi}\;(1 - 3\cos^2\theta)$ & $\sqrt\frac{5}{16\pi}\;(x^2 + y^2 - 2z^2)$\\
 $Y_{2,-1}$ & $\sqrt\frac{15}{8\pi}\;\cos\theta\;\sin\theta\;e^{-i\varphi}$ & $\sqrt\frac{15}{8\pi}\;z(x - i y)$ \\
 $Y_{2,-2}$ & $\sqrt\frac{15}{32\pi}\;\sin^2\theta\;e^{-2i\varphi}$ & $\sqrt\frac{15}{32\pi}\;(x - i y)^2$ \\
\hline
\end{tabular}
\end{center}

Let $\calY_{k m 0} = Y_{\ell m}\circ p$, with $k=2\ell$,  denote
the pullback of $Y_{\ell m}$ under the action of the Hopf
map~(\ref{HopfMap}). In accordance with
Proposition~\ref{SectionLifts}.2, its degree  is $k$. For example,
$Y_{2,0} = \sqrt\frac{5}{16\pi}\;(x^2 + y^2 - 2z^2)$ lifts to
$\calY_{4,0,0} = \sqrt\frac{5}{4\pi}\;(\alpha^2\bar\alpha^2 -
4\alpha\bar\alpha\beta\bar\beta + \beta^2\bar\beta^2)$, of degree
4.
\\

The $\calY_{k m 0}$ are simply the realization of the $Y_{\ell m}$
on the abstract 2-sphere $S$ of Clifford parallels.  As such, the
linear independence of the $Y_{\ell m}$ immediately implies the
linear independence of the $\calY_{k m 0}$ as well.

\subsection{Twist}
\label{SectionTwist}

Each $\calY_{k m 0}$ is constant along Clifford parallels, but
more general functions are not.  As we take one trip around a
Clifford parallel $(e^{it}\alpha_0, e^{it}\beta_0)$, $0 \leq t
\leq 2\pi$, the value of the monomial $\alpha^a \bar\alpha^b
\beta^c \bar\beta^d$ varies as $e^{i(a - b + c - d)}$ times the
constant $\alpha_0^{~a} \bar\alpha_0^{~b} \beta_0^{~c} \bar\beta_0^{~d}$.  In
other words, the value of a typical monomial $\alpha^a
\bar\alpha^b \beta^c \bar\beta^d$ rotates counterclockwise $(a - b
+ c - d)$ times in the complex plane as we take one trip around
any Clifford parallel.  The graph of the monomial is a helix
sitting over the Clifford parallel, motivating the following
definition.
\\

\noindent{\bf Definition \ref{SectionTwist}.1.}  The {\it twist}
of a monomial $\alpha^a \bar\alpha^b \beta^c \bar\beta^d$ is the
power of the unbarred variables minus the power of the barred
variables, i.e. $a - b + c - d$.  The {\it twist} of a polynomial
is the common twist of its terms, in cases where those twists all
agree; otherwise it is undefined.
\\

\noindent{\bf Proposition \ref{SectionTwist}.2.}  The polynomials
of well-defined twist (including all monomials) are precisely the
eigenmodes of the operator
\begin{equation}
\label{TwistMeasurement}
    \alpha\partial_\alpha - \bar\alpha\partial_{\bar\alpha}
  + \beta\partial_\beta - \bar\beta\partial_{\bar\beta},
\end{equation}
with the twist as eigenvalue.
\\

\noindent{\it Proof.}  Apply the operator to $\alpha^a
\bar\alpha^b \beta^c \bar\beta^d$ and observe the result.
$\blacksquare$\\

Geometrically, operator~(\ref{TwistMeasurement}) is essentially
the directional derivative operator in the direction of the the
Clifford parallels, the only difference being that the directional
derivative includes a factor of $i$ that
operator~(\ref{TwistMeasurement}) does not, because the
complex-valued derivative is $90^\circ$ out of phase with the
value of the function itself.

Because we consider modes of even $k$ only, the twist will always
be even.  Henceforth, for notational convenience, we shall take
our twist-measuring operator to be
\begin{equation}
\label{TwistMeasurementHalf}
  Z = \frac{1}{2}\;
    (\alpha\partial_\alpha - \bar\alpha\partial_{\bar\alpha}
   + \beta\partial_\beta - \bar\beta\partial_{\bar\beta}).
\end{equation}
The {\it ad hoc} factor of $1/2$ transforms the range of
eigenmodes from even integers to all integers.

\subsection{Siblings and the twist operators}
\label{SectionSiblings}

The twist operators
\begin{eqnarray}
\label{TwistOperators}
  twist
    \;\equiv\;
    - \bar\beta~\partial_\alpha
    + \bar\alpha~\partial_\beta \nonumber\\
  \overline{twist}
    \;\equiv\;
    - \beta~\partial_{\bar\alpha}
    + \alpha~\partial_{\bar\beta}
\end{eqnarray}
(defined in \cite {WeeksPol}) increase and decrease a function's
twist. That is, the $twist$ operator converts an $n$-eigenmode of
$Z$ to an $(n+1)$-eigenmode of $Z$, and inversely for
$\overline{twist}$. Here is the proof: It is easy to check that
the commutator $[Z,twist] = twist$, so given $ZF = \lambda F$ it
follows that
$$ Z(twist F)
  = (Z twist) F
  = (twist Z + twist)F
  = ( \lambda + 1)(twist F).$$
Thus the operator $twist$ increases by one unit the eigenvalue of
an eigenfunction of $Z$, and similarly $\overline{twist}$
decreases it by one unit.

Because $\Delta_{S^3}$ and $twist$ commute (see \cite{WeeksPol}),
the twist operator transforms each $k$-eigenmode into another
$k$-eigenmode.

Being vertical, each $\calY _{km0}$ is an eigenmode of $Z$ with
eigenvalue 0.  Repeatedly applying the operator $twist$ gives
eigenmodes of $Z$ with eigenvalues 1, 2, \ldots, $k/2$ ($k$ is
even), while repeatedly applying the operator $\overline{twist}$
gives modes with eigenvalues -1, -2, \ldots, -$k/2$.  Why do the
sequences stop at $n = \pm k/2$?  The explanation is as follows.
When written as a polynomial in the complex variables $\{\alpha,
\bar\alpha, \beta, \bar\beta\}$, the original vertical mode
$\calY_{km0}$ contains equal powers of the barred variables
$\bar\alpha$ and $\bar\beta$ and the unbarred variables $\alpha$
and $\beta$.  The operator $twist$ replaces a barred variable with
an unbarred one, keeping the degree constant while increasing the
difference $\#unbarred - \#barred$ by two.  After $\frac{k}{2}$
applications of the $twist$ operator, the polynomial
$twist^{k/2}\;\calY _{km0}$ contains unbarred variables alone: it
has maximal positive twist and further application of the $twist$
operator collapses it to zero. Analogously, the $\overline{twist}$
operator converts unbarred variables to barred ones, until
$\overline{twist}^{k/2}\;\calY_{km0}$ consists of barred variables
alone, after which further applications of $\overline{twist}$
collapse it to zero.

Let $\calY_{kmn}$ be the resulting modes. That is, for $n = 1, 2,
\ldots, k/2$, define
\begin{eqnarray}
  \calY_{k,m,+n} &=& twist^n\;\calY _{km0} \nonumber\\
  \calY_{k,m,-n} &=& \overline{twist}^n\;\calY _{km0}
\end{eqnarray}
Each $\calY_{k,m,n}$ is simultaneously a $k$-eigenmode of the
Laplacian and an $n$-eigenmode of $Z$.

\begin{table}[htdp]
\begin{center}
\begin{tabular}{ccccccccccccc}
 $\calY_{k,+\ell,-k/2}$ & $\leftarrow$ & $\cdots$ & $\leftarrow$ & $\calY_{k,+\ell,-1}$ & $\leftarrow$ & $\calY_{k,+\ell,0}$ & $\rightarrow$ & $\calY_{k,+\ell,+1}$ & $\rightarrow$ & $\cdots$ & $\rightarrow$ & $\calY_{k,+\ell,+k/2}$ \\
  & & & & & & $\Uparrow $ & & & & & & \\
  & & & & & & $Y_{\ell,+\ell}$ & & & & & & \\
 $\vdots$ & & & & $\vdots$ & & $\vdots$ & & $\vdots$ & & & & $\vdots$ \\
  & & & & & & & & & & & & \\
 $\calY_{k,+1,-k/2}$ & $\leftarrow$ & $\cdots$ & $\leftarrow$ & $\calY_{k,+1,-1}$ & $\leftarrow$ & $\calY_{k,+1,0}$ & $\rightarrow$ & $\calY_{k,+1,+1}$ & $\rightarrow$ & $\cdots$ & $\rightarrow$ & $\calY_{k,+1,+k/2}$ \\
  & & & & & & $\Uparrow $ & & & & & & \\
  & & & & & & $  Y_{\ell,+1}$ & & & & & & \\
  & & & & & & & & & & & & \\
 $\calY_{k,0,-k/2}$ & $\leftarrow$ & $\cdots$ & $\leftarrow$ & $\calY_{k,0,-1}$ & $\leftarrow$ & $\calY_{k,0,0}$ & $\rightarrow$ & $\calY_{k,0,+1}$ & $\rightarrow$ & $\cdots$ & $\rightarrow$ & $\calY_{k,0,+k/2}$ \\
  & & & & & & $\Uparrow $ & & & & & & \\
  & & & & & & $  Y_{\ell,0}$ & & & & & & \\
  & & & & & & & & & & & & \\
 $\calY_{k,-1,-k/2}$ & $\leftarrow$ & $\cdots$ & $\leftarrow$ & $\calY_{k,-1,-1}$ & $\leftarrow$ & $\calY_{k,-1,0}$ & $\rightarrow$ & $\calY_{k,-1,+1}$ & $\rightarrow$ & $\cdots$ & $\rightarrow$ & $\calY_{k,-1,+k/2}$ \\
  & & & & & & $\Uparrow $ & & & & & & \\
  & & & & & & $  Y_{\ell,-1}$ & & & & & & \\
 $\vdots$ & & & & $\vdots$ & & $\vdots$ & & $\vdots$ & & & & $\vdots$ \\
  & & & & & & & & & & & & \\
 $\calY_{k,-\ell,-k/2}$ & $\leftarrow$ & $\cdots$ & $\leftarrow$ & $\calY_{k,-\ell,-1}$ & $\leftarrow$ & $\calY_{k,-\ell,0}$ & $\rightarrow$ & $\calY_{k,-\ell,+1}$ & $\rightarrow$ & $\cdots$ & $\rightarrow$ & $\calY_{k,-\ell,+k/2}$ \\
  & & & & & & $\Uparrow $ & & & & & & \\
  & & & & & & $  Y_{\ell,-\ell}$ & & & & & & \\
\end{tabular}
\end{center}
\caption{Each $\ell$-eigenmode $Y_{\ell,m}$ of $S^2$ (\ie~each
spherical harmonic; \emph{middle column, lower entries}) lifts via
the Hopf map ($\Uparrow$) to a 0-twist $k$-eigenmode
$\calY_{k,m,0}$ of $S^3$ (\emph{middle column, upper entries}),
with $k = 2\ell$. The positive twist operator ($\rightarrow$) then
takes $\calY_{k,m,0}$ to its $\frac{k}{2}$ positively twisted
siblings (\emph{right side}) while the negative twist operator
($\leftarrow$) takes $\calY_{k,m,0}$ to its $\frac{k}{2}$
negatively twisted siblings (\emph{left side}), for a total of $(k
+ 1)^2$ linearly independent modes.}
\label{modeTable}
\end{table}

The modes $\{ \calY_{k,m,n}\}_{ n=-k/2...k/2}$ , being eigenmodes
with different eigenvalues, are linearly independent \cite
{WeeksPol}. Conclusion:  each $Y_{\ell m}$ generates, via the lift
from $S^2$ to $S^3$ (Sections \ref{SectionLifts} and
\ref{SectionLiftingModes}) and the twist operators, a
$(k+1)$-dimensional vector space $\calV ^{km\cdot }$ of $k$-modes,
with basis $\{\calY _{k, m, n}\}_{ n=-k/2...k/2}$ (see
Table~\ref{modeTable}). Thus the $2\ell +1 =k+1$ spherical
harmonics $Y_{\ell m}$ generate the complete vector space of
$k$-eigenmodes of $S^3$,
$$\calV ^{k }=\bigoplus _m  \calV ^{km\cdot},$$
with basis $\{\calY _{k, m, n}\}_{m=-k/2...k/2,\;{
n=-k/2...k/2}}$, and thus of dimension $(k+1)^2$.
\\

\noindent{\bf Proposition \ref{SectionSiblings}.1.}
$\calY_{k,m,n} = \overline{\calY_{k,-m,-n}}$.\\

\noindent{\it Proof.}  Each $\calY_{k,+m,0}$ is conjugate to the
corresponding $\calY_{k,-m,0}$ because they are lifts of the
standard 2-dimensional spherical harmonics $Y_{\ell,+m}$ and
$Y_{\ell,-m}$ which have this symmetry.  The twist
operators~(\ref{TwistOperators}) are complex conjugates of one
another by construction. Therefore when $n \geq 0$,
\begin{equation}
    \overline{\calY_{k,m,n}}
  = \overline{twist^n\;\calY_{k,m,0}}
  = \overline{twist}^n\;\calY_{k,-m,0}
  = \calY_{k,-m,-n},
\end{equation}
and similarly when $n \leq 0$.
$\blacksquare$\\

\noindent{\bf Proposition \ref{SectionSiblings}.2.}  By choosing
complex-conjugate coefficients $c_{k,m,n} =
\overline{c_{k,-m,-n}}$ one may recover the real-valued modes of
$S^3$ as
\begin{equation}
\label{RealCombination}
  c_{kmn} \; \calY_{k,m,n} + \overline{c_{kmn}} \; \calY_{k,-m,-n}
\end{equation}
In particular, whenever $m$ and $n$ are not both zero, the modes
\begin{equation}
  \phantom{i} \; \calY_{k,m,n} \; + \; \phantom{i} \; \calY_{k,-m,-n}
\end{equation}
\begin{equation}
  i \; \calY_{k,m,n} \; - \; i \; \calY_{k,-m,-n}
\end{equation}
are independent real-valued modes, analogous to cosine and sine,
respectively.
\\

\noindent{\it Proof.}  The mode~(\ref{RealCombination}) is its own
complex conjugate,
\begin{equation}
  \overline{c_{kmn} \; \calY_{k,m,n} + \overline{c_{kmn}} \; \calY_{k,-m,-n}}
  = c_{kmn} \; \calY_{k,m,n} + \overline{c_{kmn}} \; \calY_{k,-m,-n}
\end{equation}
and therefore real.
$\blacksquare$\\

\noindent{\bf Convention \ref{SectionSiblings}.3.}  For the
remainder of this article we will assume that all coefficients are
chosen in complex-conjugate pairs $c_{k,m,n} =
\overline{c_{k,-m,-n}}$ and therefore all modes are real-valued.
\\

%
\section {Eigenmodes of spherical spaces $S^3/\Gs$}
\label{SectionModesOfSphericalSpaces}
%

A spherical space is a quotient manifold $M=S^3/G$, with $G$ a
finite subgroup of SO(4). An eigenmode of $M$ with eigenvalue
$k~(k+2)$ corresponds naturally to a $k$-eigenmode of $S^3$ that
is $G$-invariant.  The set of all such modes forms a subspace
$\calV_M ^k$ of the vector space $\calV ^k$ of all $k$-eigenmodes
of $S^3$.  In the present article we focus on the case that $G$ is
a binary polyhedral group $\Gs$, because those spaces holds the
greatest interest for cosmology as well as being technically
easier.

\subsection {Vertical modes of $S^3/\Gs$ generate all modes of $S^3/\Gs$}
\label{SectionVerticalModesSuffice}

We will now show that when searching for $\Gs$-invariant
eigenmodes, we may safely restrict our attention to the vertical
ones.
\\

\noindent{\bf Proposition \ref{SectionVerticalModesSuffice}.1.}
Every $\Gs$-invariant mode of $S^3$ may be obtained from vertical
$\Gs$-invariant modes by applying the twist operators and taking a
sum.
\\

\noindent{\it Proof.} Let $F$ be an arbitrary $\Gs$-invariant mode
of $S^3$ (not necessarily vertical).  Express $F$ relative to the
basis $\calY_{kmn}$ (Table~\ref{modeTable}) as
\begin{equation}
  F = \sum_{kmn}c_{kmn}\calY_{kmn}
    = \sum_{kn}\left(\sum_{m}c_{kmn}\calY_{kmn}\right)
    = \sum_{kn}F_{kn},
\end{equation}
where $F_{kn} \equiv \sum_m c_{kmn}\calY_{kmn}$ is the component
of $F$ that is simultaneously a $k$-eigenvalue of the Laplace
operator $\Delta_{S^3}$ and an $n$-eigenvalue of the
twist-measuring operator $Z$
(Equation~(\ref{TwistMeasurementHalf})). By assumption each
element $\gamma \in \Gs$ preserves $F$.  Because $\gamma$ commutes
with both $\Delta_{S^3}$ and $Z$, it must preserve each $F_{kn}$
individually. (Unlike an arbitrary element of $SO(4)$, the
isometry $\gamma$ commutes with $Z$ because $\gamma$ takes
Clifford parallels to Clifford parallels.)  Thus each $F_{kn}$ is
$\Gs$-invariant.

Because $F_{kn}$ has constant twist, it is easily obtained by
applying the twist operator to a vertical function,
\begin{equation}
  F_{kn} = \sum_{m} c_{kmn} \calY_{kmn}
         = \sum_{m} c_{kmn} \; twist^n \calY_{km0}
         = twist^n \left(\sum_{m} c_{kmn} \calY_{km0}\right),
\end{equation}
where for negative $n$, $twist^n$ means $\overline{twist}^{|n|}$.
Because the twist operators $twist$ and $\overline{twist}$ commute
with each $\gamma$, each vertical function $\sum_{m} c_{kmn}
\calY_{km0}$ is $\Gs$-invariant, thus completing the proof.
$\blacksquare$\\

Like for $S^3$, the search for the eigenmodes of $S^3/\Gs$ reduces
to a search for the vertical ones, since each vertical
$\Gs$-invariant $k$-eigenmode generates, through the action of the
twist operators, a $(k+1)$-dimensional vector space of generic
$\Gs$-invariant $k$-eigenmodes.

\subsection {Modes of $S^2/\Gamma$ generate all vertical modes of $S^3/\Gs$}
\label{S2ModesSuffice}

Section~\ref{SectionLiftingModes} showed that the vertical modes
of $S^3$ are the pullbacks of the modes of  $S$.  Thus in a direct
geometrical sense, the modes of $S^2$ {\it are} the vertical modes
of $S^3$, and $\Gs$-invariance on $S^3$ corresponds directly to
$\Gamma$-invariance on $S^2$.
\\

\noindent{\bf Conclusion \ref{S2ModesSuffice}.1.}  The search for
$\Gs$-invariant eigenmodes of $S^3$ reduces to the search for
$\Gamma$-invariant eigenmodes of $S^2$.

%
\section{$\Gamma$-invariant eigenmodes of $S^2$}
\label{SectionS2Modes}
%

\subsection{Multipole vectors}
\label{SectionMultipoleVectors}

Consider $V^\ell$, the vector space of $\ell$-eigenmodes.
According to Maxwell's multipole vector decomposition of modes
\cite{Maxwell,Dennis,Copi,Katz,Land,mlrPoly,Weeks}, we may write
each eigenmode $f_ \ell \in V^\ell$ as
\begin{equation}
     f_\ell(x,y,z)
     = c \; r^{2\ell + 1} \;
     \nabla_{v_\ell} \cdots \nabla_{v_2} \nabla_{v_1} \;
     \frac{1}{r},
     \label{Maxwell}
\end{equation}
where $r = \sqrt{x^2 + y^2 + z^2}$ and the decomposition is well
defined up to flipping the signs of the direction vectors
$\{v_1,\dots,v_\ell\}$ and the scale factor, two at a time. The
ordering of the direction vectors is irrelevant.

Define an equivalence relation on $V^\ell$ setting two functions
$f$ and $f'$ to be equivalent whenever they are nonzero real
multiples of each other:
$$f \simeq f' \EQUIV f = c f', ~c \in \mathbb{R} - \{0\}.$$  All
the elements of each equivalence class $[f]$ share the same
decomposition~(\ref{Maxwell}) up to the choice of signs for the
direction vectors $\{v_1,\dots,v_\ell\}$ and the leading constant
$c$. Therefore each equivalence class $[f]$ is uniquely
represented by a set of directions $\{d_1,...,d_\ell\}$, where
each direction $d_i$ represents a line $\pm v_i$, with no concern
for the sign.  The set of all possible directions forms a real
projective plane $\R P^2 = S^2/{\pm Id}$.

\subsection{Invariant sets of directions}
\label{SectionInvariantDirections}

A class $[f]$ of   modes is $\Gamma$-invariant iff the
associated set $\{d_1,...,d_\ell\}$ is $\Gamma$-invariant.  Note
that although each symmetry $\gamma \in \Gamma$ is nominally a map
$\gamma: S^2 \rightarrow S^2$, its action on $\R P^2$ is well
defined.  To understand the possible classes $[f]$ of
$\Gamma$-invariant modes, we need to understand the possible
$\Gamma$-invariant sets $\{d_1,...,d_\ell\}$ of directions.

%
\section {Eigenmodes of the Poincar\'e dodecahedral space $S^3/I^*$}
\label{SectionPDSModes}
%

Let us now further restrict our attention to the Poincar\'e
dodecahedral space, because of the interest it holds in cosmology
as well as its greater technical ease.  In other words, let
$\Gamma$ be the icosahedral group $I$ comprising the 60
orientation-preserving symmetries of a regular icosahedron. We
will consider sets of directions $\{d_1,...,d_\ell\}$ that are
invariant under $I$. Because each direction $d_i$ is automatically
invariant under the antipodal map, the set $\{d_1,...,d_\ell\}$
will be invariant under the full group $I_h$ of 120 symmetries of
a regular icosahedron, reflections included.

\subsection{The orbifold}
\label{SectionOrbifold}

\begin{figure}[!ht]
\begin{center}
\includegraphics[width=15cm]{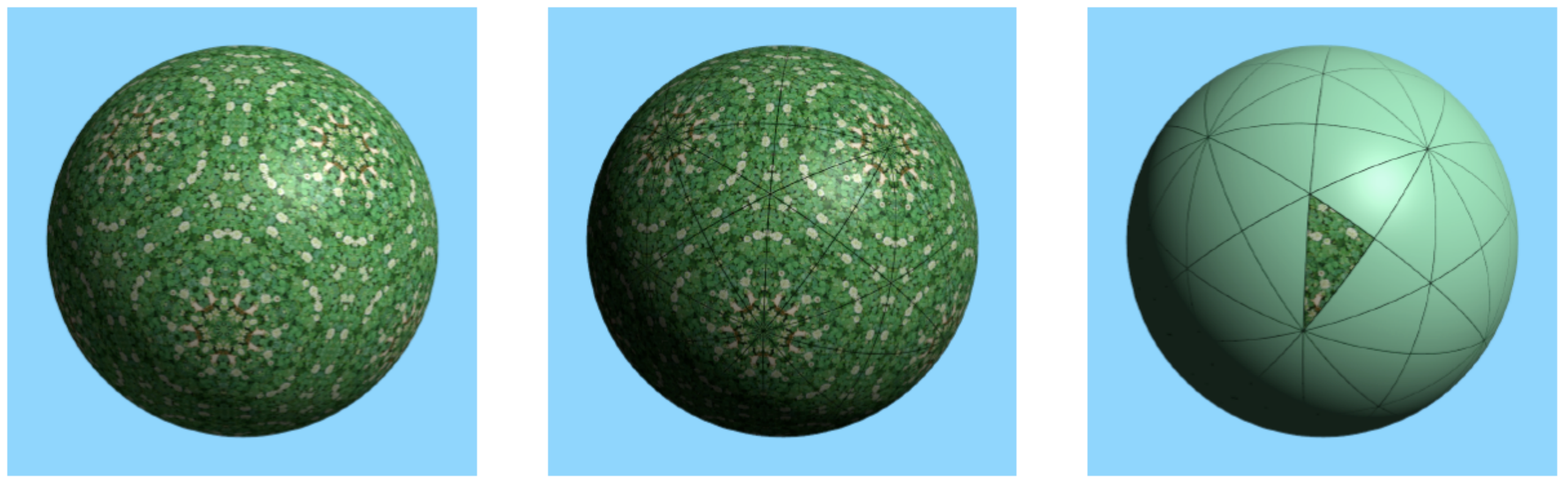}\\
\end{center}
\begin{tabular}{ccc}
(a)~~~~~~~~~~~~~~~~~~~~~~~~~~~~~~~~~~~~&(b)~~~~~~~~~~~~~~~~~~~~~~~~~~~~~~~~~~~&(c)\\
\end{tabular}
\caption{  How to construct the *235 orbifold.
     (a)  Begin with an icosahedrally symmetric pattern on the 2-sphere.
     (b)  Locate all lines of mirror symmetry.  Each is a great circle,
          and together they divide the sphere into 120 congruent
triangles.
     (c)  Fold the sphere along all mirror lines simultaneously,
          so that the whole sphere maps 120-to-1 onto a single triangle.
          The resulting quotient is the {\bf *235 orbifold}.
          The Conway notation *235 may be understood as follows:
          the `*' denotes the mirror-symmetric origin of the
triangle's sides,
          while the 2, the 3 and the 5 denote the fact that
          2, 3 and 5 mirror lines met at each corner, respectively.
} \label{orbifold}
\end{figure}

The quotient $S^2/I_h$ is an orbifold consisting of a spherical
triangle with mirror boundaries and corner reflectors with angles
$\pi/2$, $\pi/3$ and $\pi/5$ (see Figure~\ref{orbifold}). In
Conway's notation this orbifold is denoted *235.

\begin{itemize}
   \item
Each point in the \emph{interior} of the triangle lifts
to an invariant set of  120 points on $S^2$, which in turn defines an
invariant set of 60 directions.
   \item
However, each point on a \emph{mirror boundary} lifts to only 60 points
on $S^2$, defining only 30 directions.  For this reason it's
convenient to think of a point on the mirror boundary as a
``half-point''.
   \item
A point at the \emph{corner reflector} of angle $\pi/2$
lifts to 30 points on $S^2$ or 15 directions, so it's convenient
to think of it as a ``quarter point''.
\item Similarly, the points at
the corner reflectors of angle $\pi/3$ and $\pi/5$ may be
considered a $1/6$ point and a $1/10$ point, respectively.
\end{itemize}
In all cases, a $\frac{1}{F}$ fractional ($F=1,2,4,6$ or $10$)
point of   $S^2/I_h$ represents an invariant set of $\ell =
\frac{60}{F}$ directions. Another way to think about it is that a
half-point on the mirror boundary lifts to 120 half-points on
$S^2$, and then each pair of identically positioned half-points
combines to form a single full point, and similarly for the other
fractional points.\\

\noindent{\bf Definition \ref{SectionOrbifold}.1}. Let\\

$C_{\frac{1}{10}}$ denote the number of $\frac{1}{10}$ points at
the vertex of angle $\pi/5$,\\

$C_{\frac{1}{ 6}}$ denote the number of $\frac{1}{ 6}$ points at
the vertex of angle $\pi/3$,\\

$C_{\frac{1}{ 4}}$ denote the number of $\frac{1}{ 4}$ points at
the vertex of angle $\pi/2$,\\

$C_{\frac{1}{ 2}}$ denote the number of half points on the
triangle's perimeter, and\\

$C_1$ denote the number of whole points in the triangle's
interior.\\

\noindent The preceding discussion has shown that\\

\noindent{\bf Proposition \ref{SectionOrbifold}.2}.  Each
$I$-invariant equivalence class $[f]$ of modes of $S^2$
corresponds to a unique choice of $\ell$ $I$-invariant multipole
vectors. The degree of a representative mode $f$ is
\begin{equation}
\label{FractionalPoints}
  \ell \;=\;  6 C_{\frac{1}{10}}
       \;+\; 10 C_{\frac{1}{ 6}}
       \;+\; 15 C_{\frac{1}{ 4}}
       \;+\; 30 C_{\frac{1}{ 2}}
       \;+\; 60 C_1.
\end{equation}
\\

\noindent Some care is required here:  knowing that an equivalence
class $[f]$ of modes is $I$-invariant does not immediately imply
that each representative $f$ of that class is $I$-invariant.  It
is {\it a~priori} possible that some symmetry $\gamma \in I$ could
send $f$ to $-f$.  The next proposition shows that this does not
happen.
\\

\noindent{\bf Proposition \ref{SectionOrbifold}.3}.  If an
equivalence class $[f]$ of modes of $S^2$ is invariant under the
icosahedral group $I$, then each representative $f$ is also
invariant under $I$.
\\

\noindent{\it Proof.}  Let $\{d_1,...,d_\ell\}$ be the set of
$I$-invariant directions defining the class $[f]$
(Section~\ref{SectionInvariantDirections}), and let $\gamma \in I$
be a symmetry of the icosahedron. By the assumed $I$-invariance of
$[f]$, we know that $\gamma$ sends each $d_i$ to $\pm d_j$ (for
some $j$).  To prove that $f$ itself is invariant, it suffices to
prove that $\gamma$ sends $d_i$ to $-d_j$ (rather than to $+d_j$)
for an even number of the~$d_i$.

First consider the case that a given $d_i$ lies in the
``interior'' of the *235-orbifold (Figure~\ref{orbifold}c).  This
implies that 59 other $d_i$ (for different values of $i$) lie in
the interiors of other copies of the fundamental triangle
(Figure~\ref{orbifold}b), arranged symmetrically . Each
right-handed copy of the fundamental triangle lies antipodally
opposite a left-handed copy (Figure~\ref{orbifold}b).  If we make
the convention to orient each of the 60 $d_i$ in question so that
it points toward a right-handed copy of the triangle and away from
a left-handed copy, then every $\gamma \in I$ will preserve those
$d_i$ exactly, always sending a $d_i$ to a $+d_j$, never to a
$-d_j$.

Next consider the case that some $d_i$ lies on the perimeter (the
mirror boundary) of the *235-orbifold's fundamental triangle.  In
this case it has only 30 translates under the group (including
itself). The icosahedral group $I$ consists entirely of rotations,
each about some vertex of the tiling (Figure~\ref{orbifold}b). Let
$\gamma$ be some such rotation.  In the generic case that none of
the 30 $d_i$ lies exactly $90^\circ$ from the rotation axis of
$\gamma$, we may orient all 30 $d_i$ to point towards the
``northern hemisphere'' (relative to $\gamma$'s rotation axis) and
away from the ``southern hemisphere''.  In this case $\gamma$
sends each $d_i$ to a $+d_j$, never to a $-d_j$.  In the
non-generic case that some of the $d_i$ lie exactly on the
``equator'' relative to $\gamma$'s rotation axis, consider the
three sub-cases that the rotation $\gamma$ has order 2, 3 or 5.
When $\gamma$ is a rotation of order 3 or 5, easy {\it ad hoc}
conventions serve to orient the equatorial $d_i$ so that $\gamma$
respects their orientations.  When $\gamma$ is a rotation of
order~2, it perforce takes each $d_i$ to $-d_i$, but there are
exactly two such $d_i$, so the net effect is still that $\gamma$
maps the mode $f$ to $+f$, not $-f$.

Finally, consider the case that some $d_i$ lies isolated at one of
the fundamental triangle's vertices.  According to whether the
vertex is a corner reflector of order 2, 3 or 5, $d_i$ will have
15, 10 or 6 translates (including itself), respectively. Imitating
the method of the preceding paragraph, we consider a rotation
$\gamma \in I$, and wherever possible orient the $d_i$ to point
towards the northern hemisphere and away from the southern
hemisphere, thus ensuring that $\gamma$ permutes such $d_i$
respecting orientation.  It remains to consider only the $d_i$
that lie on the equator relative to $\gamma$'s rotation axis. When
$\gamma$ has order 3 or 5, its equator contains corner reflectors
of order 2 only, and an {\it ad hoc} convention serves to orient
them consistently. When $\gamma$ has order 2, it maps each
equatorial $d_i$ to $-d_i$, but the equator contains exactly four
corner reflectors of order 2, four corner reflectors of order 3
and four corner reflectors of order 5, so in each sub-case the
equator contains exactly two of the directions~$d_i$ (from among
the complete set of 15, 10 or 6 directions under consideration),
and because exactly two directions get flipped, we conclude that
$\gamma$ maps the mode $f$ to $+f$, not $-f$.
$\blacksquare$\\

\noindent{\bf Corollary \ref{SectionOrbifold}.4}.  Any value of
$\ell$ not expressible in the form (\ref{FractionalPoints}), for
example $\ell = 14$, cannot be the degree of an eigenmode of
$S^2/I$.
\\

\noindent{\bf Corollary \ref{SectionOrbifold}.5}.  The nontrivial
$I$-invariant mode of $S^2$ of least degree has degree $l = 6$.
\\

\subsection{Dimension of the space of modes}
\label{SectionDimensionOfModeSpace}

\noindent{\bf Proposition \ref{SectionDimensionOfModeSpace}.1}.
The $I$-invariant mode of degree $l =
6$ is unique up to a constant multiple.  Thus $dim(V^6) = 1$.\\

\noindent{\it Proof.}  To construct this mode, take the *235
orbifold and place a single $1/10$ point at the corner reflector
of angle $\pi/5$. This $1/10$ point lifts to 12 points of $S^2$
which in turn define 6 directions. According to Maxwell's formula
(\ref{Maxwell}), those 6 directions define an $I$-invariant class
of modes $[f]$ of degree 6.  By
Proposition~\ref{SectionOrbifold}.3, each representative $f$ of
$[f]$ is $I$-invariant.  Assuming a fixed realization of the
icosahedral group $I$, the 6 directions are well defined
--- they align with the vertices of an icosahedron or the face
centers of a dodecahedron. Therefore the class $[f]$ is also well
defined, and the only degree of freedom for the mode $f$ is the
scale factor inherent in the equivalence class $[f]$. Thus $V^6 $
is of dimension 1.
$\blacksquare$\\

The method of the preceding proposition lets us construct
$I$-invariant modes of degree 10 (place a $1/6$ point at the
corner reflector of angle $\pi/3$) and degree 15 (place a $1/4$
point at the corner reflector of angle $\pi/2$), while proving
that $I$-invariant modes of most other low degrees cannot exist.
$V^{10}$ and $V^{15}$  have dimension 1.

The case of degree 30, realized by placing a half-point on the
*235 orbifold's mirror boundary, is more interesting because we
have an extra degree of freedom corresponding to where we choose
to place the half-point.  Allowing for the scale factor inherent
in the equivalence class $[f]$ gives a total of two real degrees
of freedom: $V^{30}$ has  dimension 2.

The case of degree 60, corresponding to one full point in the *235
orbifold, is more interesting still, because now we have a choice
as to how we realize that one full point:
\begin{itemize}
   \item Case 1.  We may place a single full point anywhere
         in the orbifold.
   \item Case 2.  We may place two half-points on the orbifold's
         mirror boundary.  In the special case that the two half-points
         coincide, we get a single full point as in Case 1.
   \item Case 3.  We may place any combination of fractional points
         at the orbifold's corner reflectors, just so the fractions
         sum to one.  However it turns out that the only ways
         to do this are to place a full point at a single corner
         (for example realized as ten 1/10 points at the corner
         of angle $\pi/5$) or to place a half-point at each of two
         corners (for example realized as five 1/10 points
         at the corner of angle $\pi/5$ plus three 1/6 points
         at the corner of angle $\pi/3$).  The full point
         corresponds to Case 1 while the two half-points correspond
         to Case 2, so nothing new arises here and we will
         henceforth ignore this Case 3.
   \item Case 4.  We may place a half-point on the mirror boundary
         and a half-point's worth of fractional points at the
         corner reflectors, but as in Case 3 nothing new arises
         here so we may ignore this possibility.
\end{itemize}

\noindent{\bf Proposition \ref{SectionDimensionOfModeSpace}.2}.
The $I$-invariant classes $[f]$ of modes of $S^2$ of degree $l =
60$ are parameterized
by a real projective plane.\\

\noindent{\it Proof.}  Each class $[f]$ of degree 60 corresponds
to 60 directions $\{d_1,...,d_{60}\}$ that are invariant under the
icosahedral group $I$, which in turn correspond either to a single
point in the *235 orbifold (Case 1 above) or to a pair of
half-points on the mirror boundary (Case 2 above).

The possible locations for a whole point are obviously
parameterized by the points of the orbifold itself, which is
topologically a disk.

The possible locations for a pair of points on the orbifold's
mirror boundary are parameterized by a M\"obius strip.  To see
why, first note that the mirror boundary is topologically a circle
$S^1$.  Parameterize this circle in some arbitrary but fixed way,
with the parameter angle defined modulo $2\pi$, and then for any
pair of points define

\begin{tabular}{l}
   $\theta$ = the position of the two points' ``center of mass''
               ($\theta \in S^1 = \mathbb{R}/2\pi)$  \\
   $\phi$ = the separation between the two points
               ($\phi \in [0,\pi])$\\
\end{tabular}

\noindent At first glance this gives a cylinder parameterized by
$(\theta,\phi)$.  But $(\theta,\pi)$ and $(\theta + \pi,\pi)$
define the same pair of points, so we must identify opposite
points on the cylinder's upper boundary circle $(\theta,\pi) \sim
(\theta + \pi,\pi)$, which transforms the cylinder into a M\"obius
strip.  The cylinder's lower boundary circle $(\theta,0)$ becomes
the M\"obius strip's edge.

The M\"obius strip's edge, parameterized by $(\theta,0)$,
corresponds to the case that the two half-points fuse together to
form a single whole point on the triangle's perimeter.  This
corresponds exactly to the boundary of the disk in the whole point
parameter space.  In other words, the total parameter space is the
union of a disk and a M\"obius strip glued together along their
boundary circles, which yields a real projective plane.
$\blacksquare$\\

It's no surprise that the parameter space is a real projective
plane.  The space of $\Gamma$-invariant harmonic functions $f$ on
$S^2$ of any fixed degree $\ell$ is a vector space of some finite
dimension $n$.  When we pass from functions $f$ to equivalence
classes $[f]$ we identify each line through the origin to a single
point, giving in all cases a real projective space $\mathbb
RP^{n-1}$.  In the case just considered, with degree $\ell = 60$,
we found the projective space to be $\mathbb RP^2$ meaning the
total function space, including the scale factor, is $\mathbb
R^3$.\\

To construct a generic $I$-invariant mode, we may place any
combination of whole points (anywhere in the orbifold), half
points (on the orbifold's mirror boundary), and other fractional
points (isolated at the orbifold's corner reflectors).  Each whole
point contributes two degrees of freedom to the space of modes
(corresponding to the point's location in the 2-dimensional
triangle), each half point contributes one degree of freedom
(corresponding to its location along the triangle's 1-dimensional
perimeter), and each isolated fractional point contributes
nothing. The overall scaling factor contributes one more degree of
freedom for any nontrivial mode. In summary,
\\

\noindent{\bf Proposition \ref{SectionDimensionOfModeSpace}.3}.
The dimension of the space of $I$-invariant $\ell$-eigenmodes of
$S^2$ is given by
\begin{equation}
\label{DimensionInTermsOfC}
  dim(V^\ell) = 1 + C_{\frac{1}{ 2}} + 2 C_1.
\end{equation}
\\

Note that no matter how many half points may or may not combine
into whole points, the half and whole points together contribute
$C_{\frac{1}{ 2}} + 2 C_1$ degrees of freedom.

\subsection{Improved dimension formula}
\label{SectionImprovedDimensionFormula}

The dimension formula~(\ref{DimensionInTermsOfC}) is nice, but we
would much rather have a formula in terms of~$\ell$, to save us
the trouble of manually decomposing $\ell$ into a linear
combination of the~$C_i$.  Here is the improved formula,
\\

\noindent{\bf
Proposition~\ref{SectionImprovedDimensionFormula}.1}. The
dimension of the space of $I$-invariant $\ell$-eigenmodes of $S^2$
is given by
\begin{equation}
\label{DimensionInTermsOfEll}
  dim(V^\ell) =  1
              + \lfloor\frac{\ell}{2}\rfloor
              + \lfloor\frac{\ell}{3}\rfloor
              + \lfloor\frac{\ell}{5}\rfloor
              - \ell.
\end{equation}

\noindent{\it Proof.}  Recall that
\begin{equation}
\label{FractionalPointsBis}
  \ell \;=\;  6 C_{\frac{1}{10}}
       \;+\; 10 C_{\frac{1}{ 6}}
       \;+\; 15 C_{\frac{1}{ 4}}
       \;+\; 30 C_{\frac{1}{ 2}}
       \;+\; 60 C_1.
\end{equation}
and consider how the $C_i$ depend on $\ell$.

First consider $C_{\frac{1}{10}}$, the number of $\frac{1}{10}$
points.  Taking Equation~(\ref{FractionalPointsBis}) modulo 5 we
get $$\ell \equiv C_{\frac{1}{10}} \;\rm{(mod\,5)}.$$ But the
number of isolated $\frac{1}{10}$ points may only be 0, 1, 2, 3 or
4, because if we had 5 or more $\frac{1}{10}$ points they would
combine to form half points and acquire an additional degree of
freedom.  So the number of isolated $\frac{1}{10}$ points must be
$C_{\frac{1}{10}} = \ell - 5\lfloor\frac{\ell}{5}\rfloor$.  The
same argument, repeated mod~3 and mod~2, gives $C_{\frac{1}{6}} =
\ell - 3\lfloor\frac{\ell}{3}\rfloor$ and $C_{\frac{1}{4}} = \ell
- 2\lfloor\frac{\ell}{2}\rfloor$, respectively.

Rearranging Equation~(\ref{FractionalPointsBis}) now gives
\begin{eqnarray}
\label{RemainingDegreesOfFreedom}
  C_{\frac{1}{ 2}} + 2 C_1
  &=& \frac{1}{30}\left[
       \ell
       -  6 C_{\frac{1}{10}}
       - 10 C_{\frac{1}{ 6}}
       - 15 C_{\frac{1}{ 4}}\right]\nonumber\\
  &=& \frac{1}{30}\left[
       \ell
       -  6 \left(\ell - 5\lfloor\frac{\ell}{5}\rfloor\right)
       - 10 \left(\ell - 3\lfloor\frac{\ell}{3}\rfloor\right)
       - 15 \left(\ell - 2\lfloor\frac{\ell}{2}\rfloor\right)\right]\nonumber\\
  &=& \lfloor\frac{\ell}{5}\rfloor
    + \lfloor\frac{\ell}{3}\rfloor
    + \lfloor\frac{\ell}{2}\rfloor
    - \ell
\end{eqnarray}
Substituting Equation~(\ref{RemainingDegreesOfFreedom}) into
Equation~(\ref{DimensionInTermsOfC}) gives the final
result~(\ref{DimensionInTermsOfEll}) as stated above.
$\blacksquare$\\

This agrees with Ikeda's formula \cite{Ikeda}, while at the same
time providing a concrete construction of the modes and shedding
additional light on the formula's geometrical origins, as degrees
of freedom in an orbifold.

%
\section{Conclusion}
\label{SectionConclusion}
%

Returning to the 3-dimensional Poincar\'e dodecahedral space
$S^3/I^*$, the results of the preceding sections may be summarized
as follows.  Keep in mind that $S^3/I^*$ admits $k$-modes for even
$k$ only;  odd $k$-modes cannot exist because $I^*$ contains the
antipodal map.
\\

\noindent{\bf Theorem~\ref{SectionConclusion}.1}  To construct the
modes of the Poincar\'e dodecahedral space $S^3/I^*$,
\begin{itemize}
  \item Each mode of $S^3/I^*$ corresponds to an $I^*$-invariant
         mode of $S^3$ (elementary).
  \item Each $I^*$-invariant mode of $S^3$ is a sum of twists
         of $I^*$-invariant vertical modes of $S^3$
         (Proposition~\ref{SectionVerticalModesSuffice}.1).
  \item Each $I^*$-invariant vertical $k$-mode of $S^3$ is the
         pull-back, under the Hopf map, of an $I$-invariant
         $\ell$-mode of $S^2$, with $k = 2\ell$
         (Proposition~\ref{SectionLiftingModes}.1).
  \item The $I$-invariant $\ell$-modes of $S^2$ are parameterized
         by $\ell/60$ points on the *235-orbifold, possibly
         including fractional points
         (Section~\ref{SectionOrbifold}).
\end{itemize}

\noindent{\bf Theorem \ref{SectionConclusion}.2} The space of
$k$-modes of the Poincar\'e dodecahedral space $S^3/I^*$ has
dimension
\begin{equation}
    (k + 1) ( 1
              + \lfloor\frac{k/2}{2}\rfloor
              + \lfloor\frac{k/2}{3}\rfloor
              + \lfloor\frac{k/2}{5}\rfloor
              - \frac{k}{2}).
\end{equation}
\\

\noindent{\it Proof.}  The space of $I$-invariant $k/2$-modes of
the 2-sphere has dimension
              $ 1
              + \lfloor\frac{k/2}{2}\rfloor
              + \lfloor\frac{k/2}{3}\rfloor
              + \lfloor\frac{k/2}{5}\rfloor
              - \frac{k}{2}$
(Proposition~\ref{SectionImprovedDimensionFormula}.1) and thus the
space of vertical $I^*$-invariant $k$-modes of the 3-sphere has
this same dimension (Theorem~\ref{SectionConclusion}.1).  The
twist operators then take each vertical mode to a
$(k+1)$-dimensional space of generic $I$-invariant modes
(Table~\ref{modeTable} and
Proposition~\ref{SectionVerticalModesSuffice}.1,), completing the
proof.
$\blacksquare$\\

\end{document}